\documentclass{ws-ijmpcs}

\begin{document}

\markboth{Ulisses Barres de Almeida}
{MAGIC Extragalactic Relativistic Sources}

%
\catchline{}{}{}{}{}
%

\title{Results from MAGIC Observations of \\ 
Extragalactic Relativistic Sources}

\author{Ulisses Barres de Almeida}

\address{Max-Planck-Institut f\"{u}r Physik\\
M\"{u}nchen, D-80805,Germany\\
ulisses@mppmu.mpg.de}

\author{on behalf of the MAGIC Collaboration}

\address{}

\maketitle

\begin{history}
\received{Day Month Year}
\revised{Day Month Year}
\end{history}

\begin{abstract}
The Major Atmospheric Gamma-ray Imaging Cherenkov (MAGIC) experiment is an array of two 17-meter telescopes located in the 
Canary Island of La Palma that observes the very-high energy (VHE) gamma-ray sky in stereoscopic mode since 2009. MAGIC is
distinguished by its low-energy threshold of approximately 50 GeV, which grants the system a unique potential in the study
of distant extragalactic sources whose gamma-ray emission is significantly attenuated due to absorption by the 
extragalactic 
background light (EBL). The observation of non-thermal gamma rays in the GeV-TeV range from extragalactic sources is a 
characteristic signature of their relativistic nature and therefore fundamentally important for our understanding of the 
physics of these objects. Since the beginning of its stereo operation, MAGIC has observed a large number of active galactic
nuclei (AGN) of different classes, including several blazars and distant quasars. In this paper we will review some of the
most important results of these observations.

\end{abstract}

\ccode{PACS numbers: 95.55.Ka, 98.54.-h, 98.62.Nx}

\section{Introduction}	

Extragalactic relativistic sources such as active galactic nuclei (AGN) are among the strongest
emitters of VHE gamma-ray radiation ($E > 100$ GeV), an unavoidable signature of intense particle acceleration up to tens 
of TeVs. The study of such energetic processes is key to understanding the nature and 
evolution of these objects, as it has been revealed that gamma-ray emission accounts for a large fraction of the radiative
output from many of them. Efficient particle acceleration in the rarefied and strongly magnetised plasmas characteristic 
of relativistic jets and other astrophysical environments is \textit{per se} a question of great interest, with direct 
impact to the issue of the origin of the VHE cosmic rays\cite{Felix_rev}$^,$\cite{Jim_rev}.     

In this brief review we are going to present a selection of observational results by MAGIC which we consider to be the 
most relevant achievements in the extragalactic science programme of the collaboration in the past two years. Ground-based
gamma-ray astronomy has gone through a continuous development in the past years and in the extragalactic realm this 
coincides with a doubling of the number of detected sources (which now total over 45 objects) as well as the discovery of 
new classes of objects.

\subsection{The MAGIC Telescopes}  

Being the largest imaging atmospheric Cherenkov (IAC) instruments currently in operation, with 17-meter diameter dishes, 
the MAGIC telescopes enjoy the lowest observational threshold of all ground-based gamma-ray experiments 
($E_{\rm{Thresh}} \simeq 50$ GeV). Given the opacity effect imposed by the EBL on the propagation of VHE 
photons\cite{Dominguez11}$^,$\cite{Mazin07}, this characteristic gives the experiment a natural advantage to the study of 
the extragalactic 
universe. In fact, of the circa 20 new extragalactic objects discovered since 2009, over a third were due to observations 
by MAGIC, including one new source class and the farthest VHE AGN known\cite{Atels}.

MAGIC has passed through a major upgrade in the beginning of 2009, whereby a second telescope was added to the 
then monoscopic experiment, allowing for 3D stereoscopy imaging and reconstruction of the air showers, with a consequent 
improvement of the sensitivity of the new array across its entire observational range. The experiment currently counts with
an integral sensitivity above 250 GeV of 0.8\% of the Crab flux (after 50 hours observation). Its angular resolution is of
0.07$^\circ$ and the improved energy resolution now falls at the 15\% level\cite{Emiliano11}. Although most of the data 
presented here was taken in the new instrumental configuration, some of it contains observations with the stand-alone 
MAGIC-I telescope, which will be indicated whenever it is the case. For a view of the technical specifications of the 
MAGIC mono-system, please refer to the following publication\cite{MAGIC08}.

\section{MAGIC Extragalactic Results}

Before going into detail on a few specific sources, I would like to make a few quick remarks. The first refers to MAGIC's 
regular monitoring of two well-known blazars, Mkn 421 and Mkn 501, conducted as part of extensive MWL campaings. These 
observations, for which part of the 2009 data is already published\cite{Barres}$^,$\cite{Mkn421}$^,$\cite{Mkn501} will not
be discussed here since they will be treated by another dedicated presentation in this meeting\cite{David}. Their goal is 
to obtain an unbiased sample of the flux states of these sources to shed some light on the duty cycle of TeV-emitting AGN 
and constrain its low state, as well as to investigate the presence of spectral variability correlated with changes in the
flux state across the spectral energy distribution (SED). Another source regularly monitored by MAGIC is M 87, the 
strongest TeV emitter among the 
misaligned AGN\cite{Aharonian06}. It has recently been target of two MWL campaigns involving 
MAGIC\cite{M87a}$^,$\cite{M87b}. The most recent of these has shown the source to be variable in intra-day timescales, 
constraining the size of its emitting region to be only a few times that of the event horizon of the central supermassive 
black hole\cite{Raue11}. Several years of MAGIC monitoring of M 87 with MAGIC-I have produced a large database which 
allows to unprecendently constrain the low-state emission of the source. A low-level steady signal from M 87 was detected 
between 2005 and 2007 at the 7$\sigma$ level and shows a quiescent flux level above 300 GeV of about 
$5.4\times10^{-8}~\rm{TeV}^{-1}\rm{cm}^{-2}\rm{s}^{-1}$, with a non-variable spectrum characterised by a simple power law of
index $-2.21\pm0.21$\cite{Karsten11}. 

Finally, MAGIC regularly follows up triggers from optical high-states of blazars. Although the strict connection between 
optical and 
VHE activity from AGN is not yet clearly understood, this approach has proved extremely fruitful and since the start of 
the programme five new VHE AGN have been detected, among which the distant source S5 0716+714 (z = 0.31)\cite{MAGIC09} and 
the recently detected blazars B3 2247+381 (z=0.118; Atel 2910\cite{Atels}) and 1ES 1215+303 (z=0.13-0.24; 
Atel 3100\cite{Atels}). More recently, MAGIC has started a close collaboration with the Liverpool Telescope, to follow up 
in optical polarimetry sources detected at high-VHE states, in an effort to better understand the physics behind the TeV 
flares and jets.

\subsection{The BL Lac 3C 66A}

\begin{figure}[]
\centerline{\psfig{file=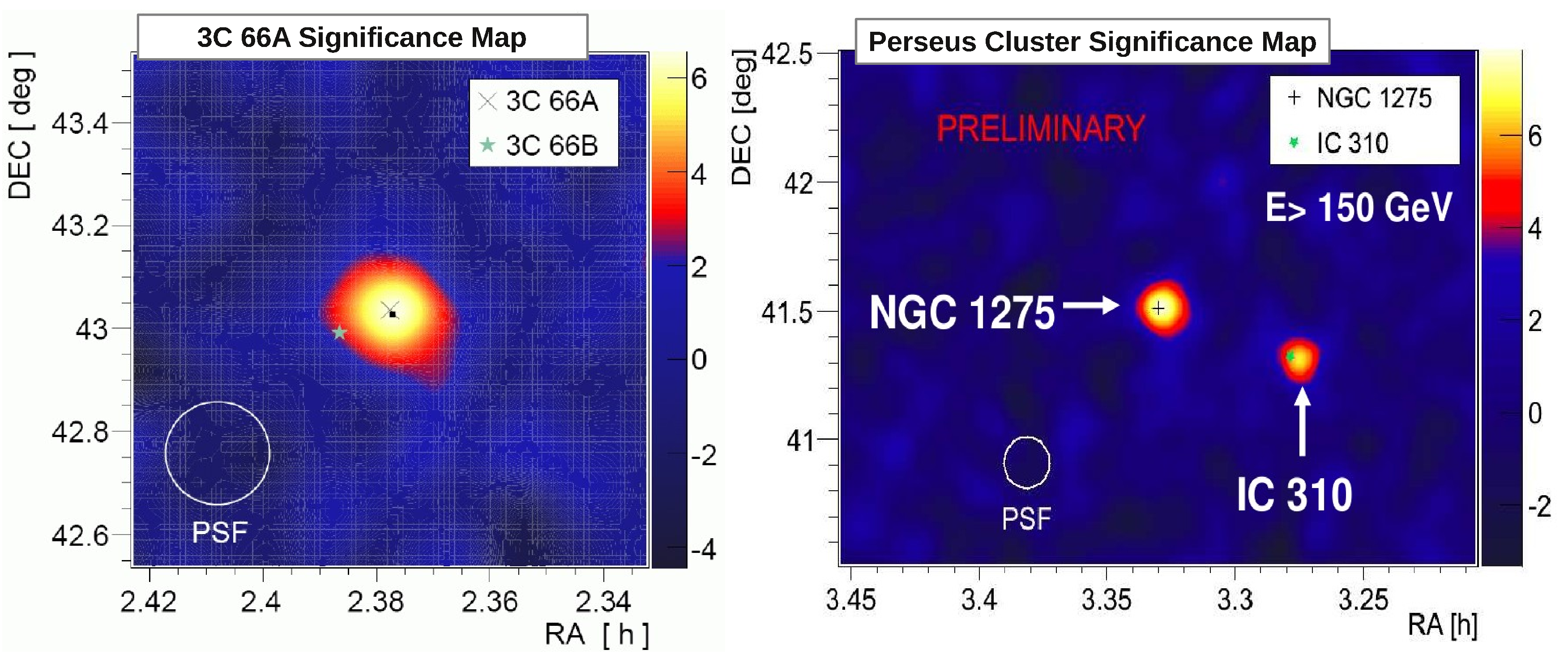,width=0.9\textwidth}}
\vspace*{8pt}
\caption{Significance maps of (\textit{left}): the blazar 3C 66A, showing also the position of the nearby radio-galaxy 
3C 66B; and (\textit{right}): the Perseus cluster of galaxies, showing emission from the central cluster galaxy, NGC 1275,
and the newly detected head-tail galaxy IC 310.
\label{fig:1}}
\end{figure}

The BL Lac 3C 66A is an example of detection which benefited from the improved angular resolution of the stereo system. 
The source, discovered by VERITAS in 2009\cite{VERITAS09}, was observed by the MAGIC telescopes between September 2009 
and January 2010, after a high-state optical trigger. Previous MAGIC-I observations from the direction of 3C 66A/B were 
not able to resolve the origin of the emission due to the proximity of a nearby radio-galaxy, 3C 66B, just 6' away
from 3C 66A\cite{MAGIC09b}. Observations at that time showed a source with harder spectrum than in the new observations 
presented here, extending up to 2 TeV, from a position compatible with 3C 66B. The achieved angular resolution of 
0.1$^\circ$ and the 
improved background rejection of the stereo system allowed nevertheless 3C 66B to be excluded as the source of the new 
measurements at the $> 4.6 \sigma$ level (Figure~\ref{fig:1}). 3C 66A was now detected\cite{MAGIC11} at a 
significance level of 6.4$\sigma$, after 2.3 hours integration time, and shows a flux of 8.3\% Crab with power law spectrum
$3.64\pm0.39_{\rm{stat}}\pm0.25_{\rm{syst}}$ between 70-500 GeV. The soft spectrum of the source has a contribution of EBL 
absorption. De-absorption conducted according to a number of different EBL models gives intrinsic spectral indexes in the 
range $\sim 2.3-2.6$, all compatible with each other. Assuming a low-level EBL and requiring the de-absorbed spectrum to be
compatible to what is observed from nearby blazars, the analysis leads to a redshift upper limit estimation of 
$z < 0.68$. This value is compatible with previous estimations of $z\sim0.444$\cite{MAGIC11} making this potentially one 
of the most distant VHE sources known.

\subsection{The Perseus Cluster}

Located at 77.7 Mpc ($z=0.018$) away, Perseus is one of the closest and largest galaxy clusters in the nearby universe. 
It is characterised by intense X-ray emission and hosts a luminous radio halo some 200 kpc across, which signs 
to the presence of diffuse Sy-emission from a high density population of cosmic-ray protons. Large inter-cluster fields, 
between 1-10~$\mu$G\cite{Vogt05}, make it an excellent enviroment for TeV particle acceleration, and therefore a 
potential source of diffuse VHE gamma-rays. 

Although no extended emission was detected, two radio galaxies were found at VHE gamma-rays from the direction of Perseus
(see Figure~\ref{fig:1}), both
of which had been detected at lower energies by the Fermi/LAT\cite{Fermi09}$^,$\cite{Neronov10}. The central cluster 
galaxy, NGC 1275\cite{MAGIC10} is distinguished by a rich filamentary structure in near-IR 
wavelengths, and strong H$\alpha$ emission, both signatures of intense interaction of the galaxy with the intracluster 
medium (ICM)\cite{Fabian08}. The galaxy was detected by MAGIC at a signifcance level of 5.2$\sigma$, with a flux of 0.03 
Crab units at $\sim$ 100 GeV and an extremely soft spectrum. The detection happened at a time of increased gamma-ray
emission in the Fermi range.
The SED of the galaxy could not be explained by an homogeneous SSC model. A structured spine-sheath jet was instead used 
to fit the spectral distribution, and the dominance of the layer's flux in the fit is in accordance with previous models 
for the emission from misaligned jets\cite{Tavecchio08}. 

IC 310 is the first exemplar of a so-called head-tail radio galaxy to be detected at VHEs. Its radio morphology is 
characterised by intense emission from the AGN (head) and an outer extension (tail), which can be interpreted as resulting from the interaction of the jet with the ICM\cite{Begelman79}.
The galaxy had been seen by the Fermi/LAT above 100 GeV\cite{Neronov10}. The MAGIC detection was achieved with a 
significance of 7.6$\sigma$ after 20 hours, and the combined Fermi+MAGIC spectrum is surprisingly hard ($\propto E^{-2}$) 
over 3 decades in energy. Despite the lack of angular resolution to directly constrain the origin of the emission, the 
indication of variability from MAGIC data suggests a compact source (the AGN) as the source of the VHE photons.
    
\subsection{Distant Sources}

\begin{figure}[]
\centerline{\psfig{file=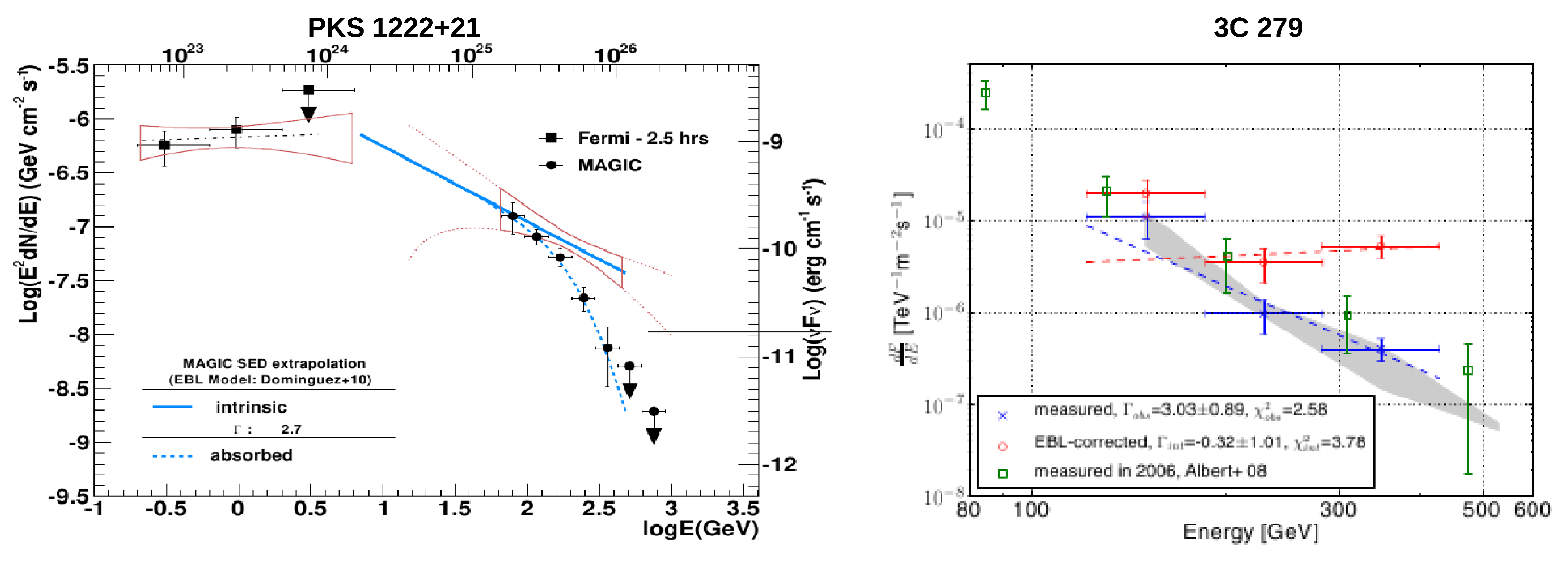,width=0.9\textwidth}}
\vspace*{8pt}
\caption{(\textit{Left}): High-energy SED component of the FSRQ PKS 1222+21 showing contemporaneous Fermi and MAGIC
measurements between 0.1 GeV and 1 TeV. (\textit{Right}): VHE spectrum of the distant quasar 3C 279 as measured by MAGIC in
2009, between 80 and 500 GeV.\label{fig:2}}
\end{figure}

Partly due to its low energy threshold, MAGIC has been responsible for the detection of several distant AGN sources, 
of which the two most distant are the FSRQs PKS 1222+21 (4C 21.35; $z = 0.432$) and 3C 279 ($z = 0.536$). 
The study of distant 
VHE-emitting AGN are of fundamental importance to constrain the EBL level\cite{Mazin07}. PKS 1222+21 ($z = 0.432$) 
was detected at a high-state in June 2010, showing short timescale variability with doubling scales shorter than 
10 min\cite{PKS1222}. The corresponding VHE spectrum of the source extends up to 500 GeV without any signs of cutoff. 
Such extended spctrum suggest that the VHE emission must come from outside the BLR to avoid internal opacity from soft 
photon fields.  
The size of the BLR for PKS 1222+21 is estimated to be of $\sim 10^{17}$ cm, implying a cross-section for the jet at the 
distance of the putative VHE emission site of at least one order of magnitude larger than the emitting region as 
constrained from the variability, thus making the case for either the presence of compact energetic embedded regions in the
jet or the occurrence of recollimation further away from the nucleus\cite{newref1}$^,$\cite{newref2}$^,$\cite{newref3}.

The quasar 3C 279, the most distant VHE source known at $z = 0.54$, was discovered by MAGIC back in 2006\cite{MAGIC08b} 
and has since been target of continual MWL monitoring, with a strong VHE flare being detected in 2007\cite{3C279}. The 
measured VHE spectrum of the source (see Figure~\ref{fig:2}), with no cut-off up to 500 GeV, has been used to put 
constraints to the EBL level, extending its probed redshift range by over a factor of 2, from $z = 0.2$ 
to $0.55$. The dataset of MAGIC on this object has been used (together with historical MWL information) to model the 
entire source's SED. Single zone leptonic (SSC+EC) models were shown to fail fitting the data, making the case for the
presence of a hadronic component to explain the high-energy end of the spectrum, with an index of about $-3.0$ up to 500 
GeV\cite{boettcher}, although a two-zone SSC+EC model was shown to also fit the data well.

\section{Conclusions}

We gave a short overview of the main extragalactic results of the MAGIC collaboration. Emphasis was given to those 
which are most interesting regarding their physical implications. Among topics left out due to space constraints is 
the status of the MAGIC searches for GRBs, for which a recent exposition can be found here\cite{GRB}.

\section*{Acknowledgments}

I would like to thank my colleagues of the MAGIC collaboration for comments and suggestions on the preparation of this 
review. The MAGIC collaboration would also like to thank the Instituto de Astrof\'{i}sica de Canarias for the excellent 
working conditions at the Observatorio Roque de los Muchachos in La Palma. We thank support from German BMBF and MPG, the 
Italian INFN adn Spanish MININN.


\end{document}